\documentclass[aps, preprint, showpacs, superscriptaddress]{revtex4-1}
\usepackage{graphicx}
\usepackage{amssymb}
\usepackage{amsfonts}
\usepackage{amsmath}
\usepackage{lmodern}
\usepackage{mathrsfs}
\usepackage{mathdots}
\usepackage{collref}
\usepackage[colorlinks=true, citecolor=blue, urlcolor=blue]{hyperref}
\begin{document}
\title{Sensitivity of parameter estimation near the exceptional point of a non-Hermitian system}
\author{Chong Chen}
 \affiliation{Department of Physics, the Centre for Quantum Coherence, and the Institute of Theoretical Physics, The Chinese University of Hong Kong, Shatin, New Territories, Hong Kong, China}
 \author{Liang Jin}
 \affiliation{School of Physics, Nankai University, Tianjin 300071, China}
\author{Ren-Bao Liu}\email{rbliu@cuhk.edu.hk}
 \affiliation{Department of Physics, the Centre for Quantum Coherence, and the Institute of Theoretical Physics, The Chinese University of Hong Kong, Shatin, New Territories, Hong Kong, China}
\date{\today}
\begin{abstract}
The exceptional points of non-Hermitian systems, where $n$ different energy eigenstates merge into an identical one, have many intriguing properties that have no counterparts in Hermitian systems. In particular, the $\epsilon^{1/n}$ dependence of the energy level splitting on a perturbative parameter $\epsilon$ near an $n$-th order exceptional point stimulates the idea of metrology with arbitrarily high sensitivity, since the susceptibility $d\epsilon^{1/n}/d\epsilon$ diverges at the exceptional point. Here we theoretically study the sensitivity of parameter estimation near the exceptional points, using the exact formalism of quantum Fisher information. The quantum Fisher information formalism allows the highest sensitivity to be determined without specifying a specific measurement approach.  We find that the exceptional point bears no dramatic enhancement of the sensitivity. Instead, the coalescence of the eigenstates exactly counteracts the eigenvalue susceptibility divergence and makes the sensitivity a smooth function of the perturbative parameter.

\end{abstract}
\maketitle

\section{Introduction}
Extending physical parameters from the real axis to the complex plane largely deepens our understanding of quantum mechanics \cite{Bender1998, Moiseyev2011} and enriches our controllability of quantum systems \cite{Mostafazadeh2009, Chong2010, Wan2011, Feng2012, Peng2014, Hodaei2014, Miao2016, Feng2017, El-Ganainy2018}. One intriguing phenomenon that emerges from this extension is the non-Hermitian degeneracy, known as the exceptional point (EP). In contrast to level degeneracy points in Hermitian systems, the EP is associated with level coalescence, in which not only the eigenenergies but also the eigenstates become identical \cite{Heiss2004, Berry2004}. Many distinctive effects without Hermitian counterparts arise around the EP, such as the square root frequency dependence \cite{Peng2014} and the nontrivial topological property  resulting from the Riemann sheet structures of the EP-ended branch-cut in the complex parameter plane \cite{Rotter2015, Doppler2016, Xu2016, Hassan2017, Leykam2017, zhou2018, Shen2018, Zhang2018, Ding2018, GoldZak2018}. Other intriguing phenomena include unidirectional reflectionless and coherent perfect absorption due to the spectral singularity in non-Hermitian systems \cite{Chong2010, Wan2011, Lin2011, Feng2012, Sun2014, Jin2018}.

Around the $n$-th order EP \cite{Heiss2008,Demange2012}, where the coalescence of $n$ levels occurs,  the eigenenergy shows an $\epsilon^{1/n}$ dependence on the perturbative parameter $\epsilon$. This result stands in sharp contrast to the Hermitian degeneracy, where the eigenenergy has a linear or high-order dependence.  That means the eigenenergies around EPs have diverging susceptibility on the parameter change since $d\epsilon^{1/n}/d\epsilon=\epsilon^{1/n-1}$ diverges at $\epsilon=0$. Based on this divergence, schemes of parameter estimation (or sensing) working around EPs were proposed for the purpose of beating the metrology limit of Hermitian systems \cite{Wiersig2014, Wiersig2016}. Recently, this idea  has been experimentally studied \cite{Chen2017, Hodaei2017, Zhao2018}. However, the diverging eigenvalue susceptibility does not necessarily lead to arbitrary high {\em sensitivity}. In parameter estimation the sensitivity is usually defined as minimum parameter change that can be determined above the noise level within a given data acquisition time. Thus defined sensitivity is more relevant than the eigenvalue susceptibility is to practical applications of parameter estimation.  In Hermitian systems, the sensitivity is inversely proportional to the eigenvalue susceptibility, i.e., the larger the susceptibility, the higher the sensitivity. Such a relation is based on the fact that all the eigenstates are distinguishable and the transitions between these eigenenergies can be excited to measure the eigenvalue susceptibility.  However, non-Hermitian systems are fundamentally different.  Because different eigenstates of non-Hermitian systems are in general non-orthogonal and even become identical at the EP, exciting the transitions between different eigenstates near the EP to measure the eigenvalue susceptibility is infeasible.

In this paper, we study the sensitivity around the EP of a coupled cavity system for its immediate relevance to recent experimental studies \cite{Chen2017, Hodaei2017}. Nonetheless, the theoretical formalism and the main conclusion - no dramatic sensitivity enhancement at the EP - are applicable to a broad range of systems, such as magnon-cavity systems \cite{Zhang2016, Zhang2017} and opto-mechanical systems \cite{Jing2014, lue2015, Xu2016a}. We use the exact formalism of quantum Fisher information (QFI) \cite{Paris2009} to  characterize the sensitivity of parameter estimation. The QFI formalism enables us to evaluate the sensitivity without referring to a specific measurement scheme - be it phase, intensity, or any other complicated measurements of the output from the system. We find that no sensitivity boost exists at the EP. The reason boils down to the coalescence of the eigenstates around the EP. Due to the indistinguishability of different eigenstates around the EP, not one but all eigenstates are equally excited by an arbitrary detection field. The average of all eigenstates exactly cancels out the singularity in the susceptibility divergence of the eigenenergies and makes the sensitivity normal around the EP.

\section{Model}
We consider two near resonance coupled cavities with the effective non-Hermitian Hamiltonian
\begin{equation}
	\hat{H}_{\mathrm{eff}}= (\nu_{a}-i \frac{\gamma_{a}}{2}) \hat{a}^{\dagger} \hat{a} + (\nu_{b}-i\frac{\gamma_{b}}{2}) \hat{b}^{\dagger} \hat{b} +g (\hat{a}^{\dagger}\hat{b}+\hat{b}^{\dagger} \hat{a}),
	\label{eq:EffectiveH}
\end{equation}
where $\nu_{a(b)}$ is the cavity frequency and $\gamma_{a(b)}$ is the decay rate induced by the photon leakage of the cavity $a(b)$, $g$ is the coupling strength, and the Planck constant $\hbar$ is taken as unity throughout this paper. For the quadratic Hamiltonian in Eq. \eqref{eq:EffectiveH}, the dynamics are captured by the coefficient matrix
\begin{equation}
	M=(\bar{\nu}-i \frac{\bar{\gamma}}{2})\mathbb{I}+\left(
	\begin{array}{cc}
		\frac{\epsilon}{2}-i\frac{\gamma}{2} & g \\
		g & -\frac{\epsilon}{2}+ i\frac{ \gamma}{2}
	\end{array}
	\right),
	\label{eq:CoefficientM}
\end{equation}
where $\bar{\nu}=\frac{\nu_{a}+\nu_{b}}{2}$ and $\epsilon=\nu_{a}-\nu_{b}$ are the average and detuning of the cavity frequencies, respectively, and $\bar{\gamma}=\frac{\gamma_{a}+\gamma_{b}}{2}$ and $ \gamma=\frac{\gamma_{a}-\gamma_{b}}{2}$ are the average and difference in decay rates, respectively. In sensing experiments, the detuning $\epsilon \rightarrow 0$ is a perturbation term and can be introduced, e.g., by a nanoparticle that changes the effective volume and hence the frequency of one of the cavities, say, cavity $a$ \cite{Chen2017}.

The eigenvalues and the corresponding right eigenvectors are obtained by diagonalizing the coefficient matrix $M$. Results are $\nu_{\pm}=\bar{\nu}-i\frac{\bar{\gamma}}{2}\pm \sqrt{g^{2}+\left(\frac{\epsilon}{2}-i\frac{\gamma}{2}\right)^2}$ and $\psi^{\mathrm{R}}_{\pm}=z_{\pm} [(\frac{\epsilon}{2}-i\frac{\gamma}{2}) \pm \sqrt{g^{2}+(\frac{\epsilon}{2}-i\frac{\gamma}{2})^{2}}, g]^{T}$, where $z_{\pm}$ are the normalization factors such that $ \psi^{\mathrm{R} \dagger}_{\pm}\psi^{\mathrm{R}}_{\pm}=1$. The left eigenvectors $\psi^{\mathrm{L}}_{\pm}=\frac{1}{z_{\pm}\sqrt{g^{2}+(\frac{\epsilon}{2}-i\frac{\gamma}{2})^{2}}}[\pm g, \sqrt{g^{2}+(\frac{\epsilon}{2}-i\frac{\gamma}{2})^{2}}\mp (\frac{\epsilon}{2}-i\frac{\gamma}{2})]$, which are in general not the Hermitian conjugate of the right eigenvectors, are determined by the conditions that $\psi^{\mathrm{L}}_{i}\psi^{\mathrm{R}}_{j}=\delta_{i,j}$. The EP occurs at $\epsilon=0$ and $g=|\gamma|/2$, where the eigenvalues are degenerate and the eigenstates coalesce. Around the EP, the energy splitting shows a square root perturbation dependence on $\epsilon$ as $\Delta\equiv(\nu_{+}-\nu_{-}) \approx 2\sqrt{|\gamma|\left(g-\frac{|\gamma|}{2}\right)-i\gamma \frac{\epsilon}{2}}$. The susceptibility of the energy splitting diverges at the EP as
\begin{equation}
	\chi\equiv \frac{\partial \Delta }{\partial \epsilon} \approx \frac{-i \gamma/2}{\sqrt{|\gamma|\left(g-\frac{|\gamma|}{2}\right)-i\gamma \frac{\epsilon}{2}}}.
	\label{eq:Susceptibility}
\end{equation}
The eigenvectors $\psi^{R}_{\pm}$ of the non-Hermitian $M$ are in general non-orthogonal and coalescent at the EP as $|\psi^{R \dagger}_{+}\psi^{R}_{-}|\approx 1-\frac{2}{|\gamma|}\sqrt{(g-\frac{|\gamma|}{2})^2+(\frac{\epsilon}{2})^{2}}$.

\section{Quantum Fisher information}

In general, the sensing can be viewed as a scattering process. The input state $\rho^{\mathrm{in}}$ after scattering with the sensing system yields an output state $\rho(\epsilon)$, which depends on the parameter $\epsilon$ that is to be estimated. Certain measurement of the output state $\rho(\epsilon)$ determines the parameter $\epsilon$. The sensitivity is defined as
\begin{equation}
	\eta=\delta \epsilon_{\mathrm{min}} \sqrt{T},
	\label{eq:Sensitivity}
\end{equation}
where $\delta \epsilon_{\mathrm{min}}$ is the minimum detectable parameter change for a detection time $T$  \cite{Degen2017}. In general, the sensitivity depends on the specific measurement scheme, which, in optics, is usually the measurement of the phase, the intensity, or various quadratures. However, there is a theoretical lower bound for all kinds of measurement, which is known as the quantum Cram\'{e}r-Rao bound \cite{Cramer1946}
\begin{equation}
	\eta \ge 1/\sqrt{F^{\epsilon} n/T}.
	\label{eq:CramerRaoBound}
\end{equation}
Here $F^{\epsilon}$ is the QFI of the output state $\rho(\epsilon)$ and $n/T$ is the number of experiment repetitions per unit time. Mathematically, QFI is defined as the infinitesimal Bures distance  between two close-by output states $\rho(\epsilon)$ and $\rho(\epsilon+\delta\epsilon)$ \cite{Braunstein1994}, namely
\begin{equation}
	F^{\epsilon}= \lim_{\delta \epsilon \rightarrow 0}\frac{4}{\delta\epsilon^{2}}d^{2}_{B}[\rho(\epsilon),\rho(\epsilon+\delta\epsilon)].
	\label{eq:QFI}
\end{equation}
Here $d_{\mathrm{B}}(\rho,\rho^{\prime})$ is the Bures distance, which describes the indistinguishability between the states $\rho$ and $\rho^{\prime}$ \cite{Uhlmann1976, Bures1969}. Formally, it has an expression
\begin{equation}
d^{2}_{\mathrm{B}}\left(\rho,\rho^{\prime}\right)= 2-2\sqrt{\mathcal{F}(\rho,\rho^{\prime})},
	\label{eq:BuresDistance}
\end{equation}
where $\mathcal{F}(\rho,\rho^{\prime})=\left[\mathrm{Tr}\sqrt{\rho^{1/2} \rho^{\prime} \rho^{1/2}}\right]^{2}$ is the fidelity between the states $\rho$ and $\rho^{\prime}$. A particular advantage of the QFI is that it is independent of the specific measurement scheme. In the following, we use the QFI to characterize the sensitivity of a non-Hermitian system. According to the definition of QFI in Eqs. \eqref{eq:QFI} and \eqref{eq:BuresDistance}, the highest sensitivity is determined by the change of the state $\rho(\epsilon)$ in response to the variation of the parameter $\epsilon$.

The output state $\rho (\epsilon)$ and the input state $\rho^{\mathrm{in}}$ are connected via the scattering process \cite{Taylor2006, Newton2013}. The input $\nu$-frequency  photon $\hat{c}^{\mathrm{in}}_{\nu}$ after scattering  by the sensing system gives the output photon $\hat{c}^{\mathrm{out}}_{\nu}$. In formula, we have
\begin{equation}
	\hat{c}^{\mathrm{out}}_{\nu}=\hat{c}_{\nu}^{\mathrm{in}}-\hat{s}^{\mathrm{in}}_{\nu},~ \hat{s}=\frac{1}{\nu+\hat{H}\wedge} [\hat{V}, \hat{c}_{\nu}],
	\label{eq:InputOutput-S}
\end{equation}
where the input and output operators are defined as $\hat{o}^{\mathrm{in/out}}(t)= \hat{\Omega}^{\dagger}_{\pm} \hat{o}(t) \hat{\Omega}_{\pm}$ with the Moller operators $\hat{\Omega}_{\pm}=\lim_{t^{\prime}\rightarrow \mp \infty} e^{i \hat{H}t^{\prime}} e^{-i \hat{H}_{0}t^{\prime}}$ and $\hat{H}=\hat{H}_{0}+\hat{V}$ is the total Hamiltonian with $\hat{H}_{0}$ being the free Hamiltonian and $\hat{V}$ being the interaction between the sensing system and the input photons (see Appendix \ref{sec:QST-QFI-app} for details). Here the symbol ``$\wedge$'' denotes the commutation operation, i.e., $\hat{A}\wedge \hat{B}=[\hat{A},\hat{B}]$ and the subscript $\nu$ in $\hat{s}^{\mathrm{in}}_{\nu}=\int \frac{dt}{\sqrt{2\pi}}e^{i\nu t} \hat{s}^{\mathrm{in}} (t)$ denotes the $\nu$-frequency component.

\begin{figure}[tpb]
	\centering
	\includegraphics[width=1.0\columnwidth]{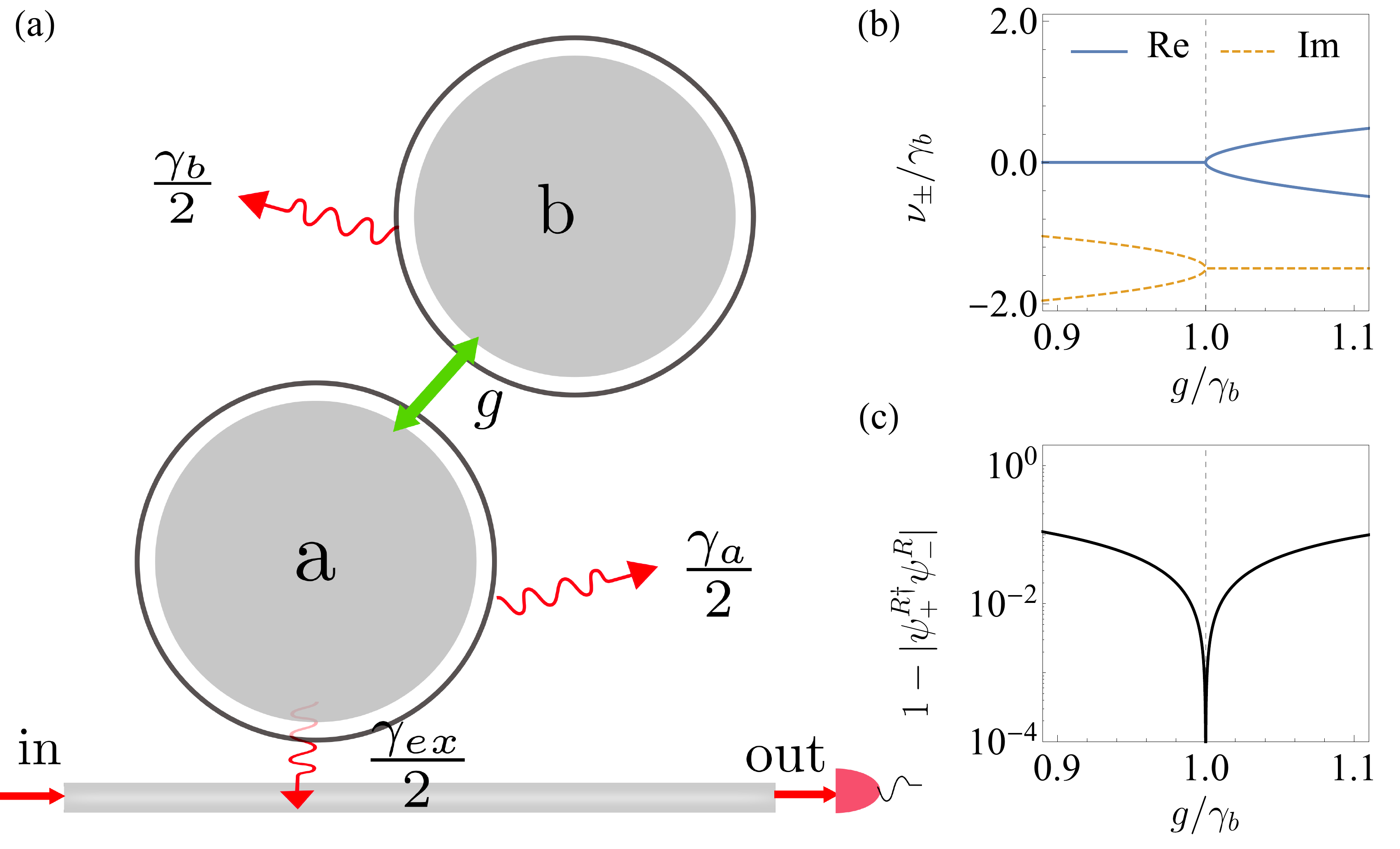}
	\caption{(a) Schematic of a coupled cavity sensing system. Two cavities, $a$ and $b$, coupled through the photon transmission with coupling strength $g$, have rates $\gamma_{a}$ and $\gamma_{b}$ of leakage to free space, respectively. A waveguide is coupled to cavity $a$, for photon input and output. The waveguide-cavity coupling strength is characterized by the decay rate $\gamma_{ex}$. (b) Energy diagram of the sensing system. The EP occurs at the point (vertical dashed line) where both the real part and the imaginary part of the eigen frequencies are degenerate. (c) Overlap of the quasinormal modes $\psi^{R}_{\pm}$. Quasinormal modes are in general non-orthogonal and coalesce at the EP. Parameters are $\gamma_{b}=1.0$, $\gamma_{a}=5.0\gamma_{b}$, $\gamma_{ex}=0.1 \gamma_{b}$, and $\nu_{a}=\nu_{b}$.}
	\label{fig:SensingScheme}
\end{figure}

We consider a general case of linear systems. The Hamiltonian reads $\hat{H}_{0}=\int d\nu \nu \hat{c}^{\dagger}_{\nu} \hat{c}_{\nu}+\sum_{lk}\hat{o}^{\dagger}_{l}M_{lk}\hat{o}_{k}$ and $\hat{V}=\int \frac{d \nu}{\sqrt{2\pi}} \sum_{j} \sqrt{\gamma_{ex,j}} (\hat{o}^{\dagger}_{j} \hat{c}_{\nu}+h.c.)$, where $[\hat{o}^\dagger_{l}, \hat{o}_{k}]=\delta_{l,k}$ are linear operators and $\gamma_{ex,j}$ is the coupling strength between the input photons and $j$-th mode of the sensing system.  For example $\hat{o}_{1}=\hat{a}$, $\hat{o}_{2}=\hat{b}$ and $\gamma_{ex,j}=\gamma_{ex}\delta_{j, 1}$ for the coupled cavity system shown in Fig. \ref{fig:SensingScheme}. Taking the interaction $\hat{V}$ as a perturbation and expanding it to the second order, we obtain
\begin{equation}
	\hat{c}^{\mathrm{out}}_{\nu}\approx \hat{c}^{\mathrm{in}}_{\nu}+\sum_{lj}(M_{\nu}^{-1})_{lj}\sqrt{\gamma_{ex,j}} [\frac{\hat{o}^{\mathrm{in}}_{l}(\nu)}{\sqrt{2\pi}}  -i\sqrt{\gamma_{ex,l}}\hat{c}^{\mathrm{in}}_{\nu}],
	\label{eq:ScatteringTheory}
\end{equation}
where $M_{\nu}=(\nu \mathbb{I}-M$) is the frequency-shifted coefficient matrix. The output state $\rho(\epsilon)=\bigotimes_{\nu} P^{\nu}_{n,m}\frac{(\hat{c}^{\mathrm{out} \dagger}_{\nu})^{n}}{\sqrt{n!}}|0\rangle\langle0| \frac{(\hat{c}^{\mathrm{out}}_{\nu})^{m}}{\sqrt{m!}}$, where $P^{\nu}_{n,m}=\langle n_{\nu}| \rho^{\mathrm{in}}|m_{\nu}\rangle$ is the density matrix element of the input state. Here we assume that the input state is a product state of different frequency modes. A small disturbance $\delta \epsilon$ of the sensing system changes the output state to
\begin{eqnarray}
	\rho(\epsilon+\delta \epsilon)&=&\rho(\epsilon)+\partial_{\epsilon}\rho(\epsilon)\delta \epsilon+\hat{O}(\delta \epsilon^2),
	\label{eq:FunctionalAnalysis}\\
	\partial_{\epsilon}\rho(\epsilon)&=& \int \frac{d \nu}{\sqrt{2\pi}} \sum_{lj}\frac{\partial \rho(\epsilon)}{\partial (M^{-1}_{\nu})_{lj}}\frac{d (M^{-1}_{\nu})_{lj}}{d \epsilon} ,
	\label{eq:Differential}
\end{eqnarray}
The QFI, with the expansion in Eq. \eqref{eq:FunctionalAnalysis} kept to the leading order of $\delta \epsilon$, becomes
\begin{equation}
	F^{\epsilon}=2\sum_{\alpha,\beta} \frac{|\langle \mu_{\alpha}| \partial_{\epsilon}\rho(\epsilon)|\mu_{\beta}\rangle|^{2}}{p_{\alpha}+p_{\beta}} ,
\label{eq:QFI-LRep}
\end{equation}
where $|\mu_{\alpha}\rangle$ is $\alpha$-th eigenstate of $\rho(\epsilon)$ with eigenvalue $p_{\alpha}$.  The output state $\rho(\epsilon)$ and its differential $\partial_{\epsilon} \rho(\epsilon)$, as functions of $(M^{-1}_{\nu})_{lj}$, are well defined unless the matrix $M_{\nu}$ is singular, i.e., $\det[M_{\nu}]=0$ and $M^{-1}_{\nu}$ is divergent. Note that such a singular condition is independent of the EP. For example, the coefficient matrix in Eq. \eqref{eq:CoefficientM} shows no divergence at the EP as $\det[M_{\nu}]=(\nu-\bar{\nu}+i \frac{\bar{\gamma}}{2})^{2}\neq 0$ for all frequencies. Therefore, the QFI for a sensing system with well defined $M_{\nu}$ shows no $\epsilon$ singularity at the EP.

\begin{figure}[tpb]
	\centering
	\includegraphics[width=1.0\columnwidth]{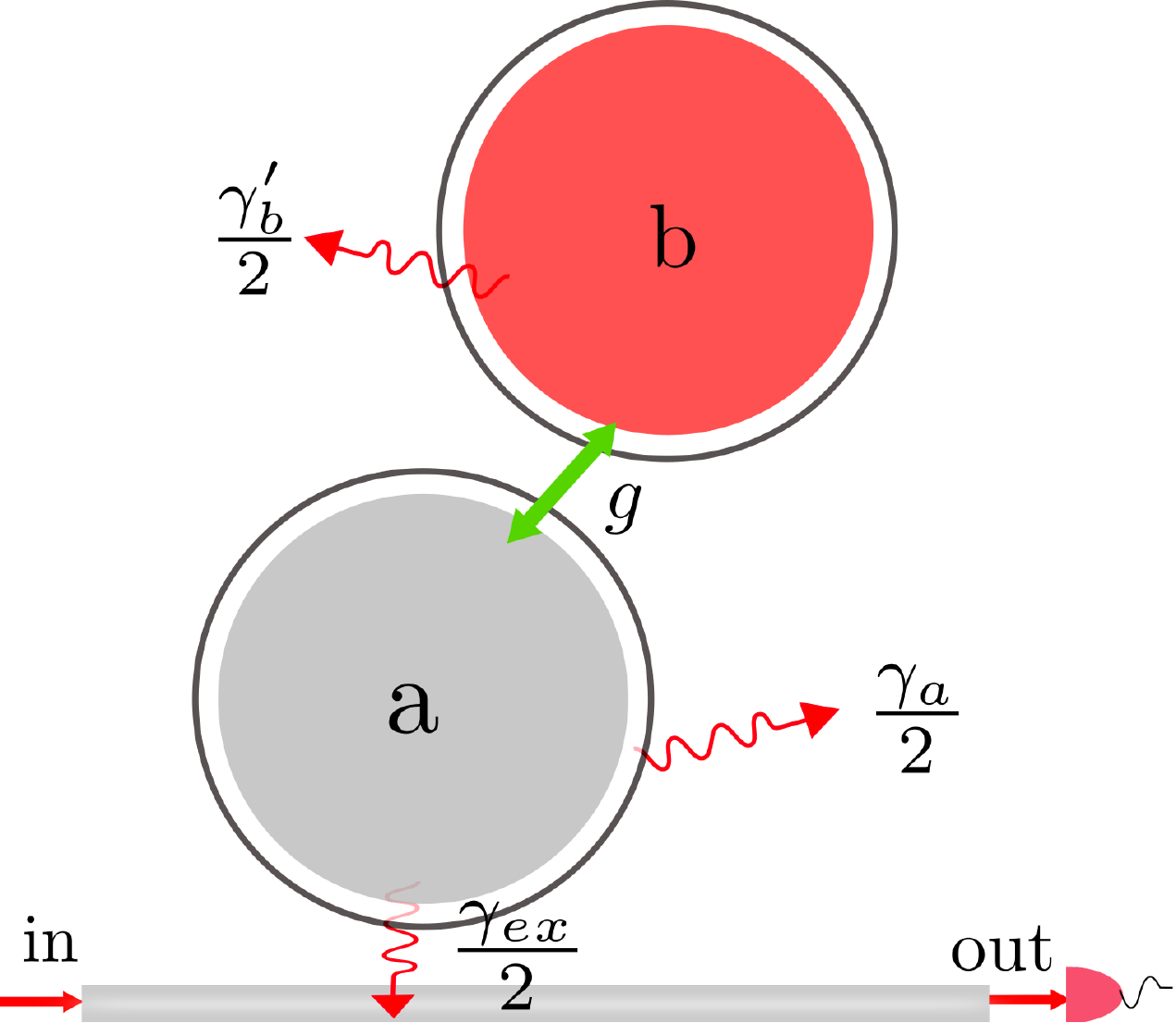}
	\caption{The active-passive coupled cavity sensing system. Here cavity $b$ is filled with a gain medium. The effective decay rate $\gamma^{\prime}_{b}$ is tuned by varying the optical pumping power.}
	\label{fig:SensingScheme-G}
\end{figure}

For the completeness of discussion, we briefly comment on sensing systems with $\det[M_{\nu}]=0$. In such a case, $\rho(\epsilon)$ and its differential $\partial_{\epsilon}\rho(\epsilon)$, in general, are singular because the divergence of $(M^{-1}_{\nu})_{lj}$ makes the output state $\rho(\epsilon)$ sensitive to the parameter $\epsilon$. A small change of $\epsilon$ can make an abrupt change of $\rho(\epsilon)$. In physics, the abrupt change of the output state indicates a non-equilibrium phase transition. An explicit example is the lasing transition of a gain cavity system \cite{DeGiorgio1970}. By embedding a gain medium into cavity $b$ and applying optical pumping, the effective decay rate is effectively reduced to $\gamma_{b}^{\prime}$ and even change its sign (see Fig. \ref{fig:SensingScheme-G}). That yields the lasing threshold $\gamma^{\prime}_{b}=-4\frac{g^2}{\gamma_{a}}$ if $g<\gamma_{a}/2$ or $\gamma^{\prime}_{b}=-\gamma_{a}$ if $g\ge \gamma_{a}/2$. Above the threshold, the system is in lasing phase. The singular point is in general not related to the EP that occurs at $g=|\gamma|/2=(\gamma_{a}-\gamma_{b}^{\prime})/4$, unless the non-equilibrium phase transition coincides with the EP. An example is the $\mathcal{PT}$ phase transition that occurs at $g=\gamma_{a}/2$ and $\gamma^{\prime}_{b}=-\gamma_{a}$. But even for such coincidence, the divergence of QFI is caused by the phase transition rather than the EP. This is evidenced by the fact that $F^{\epsilon}$, as a function of $(M^{-1}_{\nu})_{lj}$, diverges as $\int \frac{d \nu}{2\pi}|(M^{-1}_{\nu})_{lj}|^{\alpha}$ near the transition point, where $\alpha$ is the critical exponent. For example, $\alpha=2$ for the lasing transition (See  Appendix \ref{sec:APsystem-app} for details). The above discussions are based on a linear theory in which the dynamics of the sensing system are captured by a linear matrix $M_{lk}$.  However, near the  non-equilibrium transition, the critical fluctuations diverge and their effects become nonlinear. The diverging critical fluctuations may prevent the singular behavior of the QFI. A systematic study on the competition between the critical fluctuation and the abrupt change of output near the non-equilibrium phase transition is needed before a conclusion can be made on whether the phase transition can dramatically enhance the sensitivity of parameter estimation, which, however, is beyond the scope of this paper.


Physically, the lack of divergence of QFI at the EP is due to the coalescence of the eigenvectors (quasinormal modes). The coefficient matrix in Eq. \eqref{eq:CoefficientM} can be diagonalized as $M_{\nu}=VD_{\nu}V^{-1}$, where $D_{\nu}$ is a diagonal matrix of eigenvalues $(\nu-\nu_{\pm})$ and $V$ is the matrix composed of the eigenvectors $\psi^{R}_{\pm}$. The differential is $\frac{d M_{\nu}}{d \epsilon}=\frac{1}{2}(\mathbb{I}+\sigma_{z})+\frac{1}{2}[(\frac{\psi^{R}_{+,1}}{\psi^{R}_{+,2}}+\frac{\psi^{R}_{-,1}}{\psi^{R}_{-,2}})+(\frac{\psi^{R}_{+,1}}{\psi^{R}_{+,2}}-\frac{\psi^{R}_{-,1}}{\psi^{R}_{-,2}})\chi]\sigma_{+} $, where $\sigma_{x/y/z}$ are the Pauli matrices,  $\sigma_{\pm}=\frac{1}{2}(\sigma_{x}\pm i \sigma_{y})$, and $\psi^{R}_{\pm,i}$ denotes the $i$-th element of the right eigenvector $\psi^{R}_{\pm}$.  The term $(\frac{\psi^{R}_{+,1}}{\psi^{R}_{+,2}}-\frac{\psi^{R}_{-,1}}{\psi^{R}_{-,2}})=\frac{\Delta}{g}$ vanishes at the EP due to the eigenvector coalescence, canceling the $\Delta^{-1}$ susceptibility divergence near the EP.

The analysis based on Eq. \eqref{eq:FunctionalAnalysis} is applicable for a coefficient matrix $M_{\nu}$ of any dimensions and hence an EP of arbitrary order. Therefore, the QFI shows no divergence at the EP in general.

\section{Input-output theory}
We consider the configuration of a coupled cavity system with input and output channels (as shown in Fig. \ref{fig:SensingScheme}). The QFI is extracted from the output for the parameter estimation (e.g., estimation of the frequency of a cavity).  In addition to the waveguide input and output, we also include the realistic leakage into the free space, with rates $\gamma_{a/b}$ for cavity $a/b$. The Hamiltonian of the open system is written as
\begin{equation}
\hat{H}=\hat{H}_{S}+\hat{H}_{I}+\hat{H}_{B},
\label{eq:TotalHamiltonian}
\end{equation}
where $\hat{H}_{S}=\nu_{a} \hat{a}^{\dagger} \hat{a} + \nu_{b} \hat{b}^{\dagger} \hat{b} + g (\hat{a}^{\dagger} \hat{b} + \hat{b}^{\dagger} \hat{a})$ is the Hamiltonian of the coupled cavity system, $\hat{H}_{I}=  \int \frac{d\nu}{\sqrt{2\pi}} [ \hat{a}^{\dagger} (\sqrt{\gamma_{a}} \hat{a}_{\nu}+\sqrt{\gamma_{ex}} \hat{c}_{\nu})+\sqrt{\gamma_{b}} \hat{b}^{\dagger} \hat{b}_{\nu}+h.c.]$ is the coupling to the open channel and the free space photons, and $\hat{H}_{B}=\int d\nu \nu(\hat{a}^{\dagger}_{\nu} \hat{a}_{\nu}+\hat{b}^{\dagger}_{\nu} \hat{b}_{\nu}+\hat{c}^{\dagger}_{\nu} \hat{c}_{\nu})$ is the non-interacting Hamiltonian of the open channel and the free space photons. Here $\hat{a}_{\nu} (\hat{b}_{\nu})$ and $\hat{c}_{\nu}$ are the $\nu$ frequency annihilation operators of the free space photons for cavity $a$ ($b$) and the waveguide photons, respectively.
For the input state (the photon state in the remote past $t=-\infty$) $\rho^{\mathrm{in}}:=\rho_{a}(-\infty)\otimes \rho_{b}(-\infty)\otimes \rho_{c}(-\infty)$, we are to determine the waveguide output state at the remote future $\rho^{\mathrm{out}}_{c}:=\rho_{c}(\infty)$.
In the Markovian noise process, the input-output theory gives \cite{Collett1984}
\begin{equation}
	\hat{c}^{\mathrm{out}}(t)-\hat{c}^{\mathrm{in}}(t)=-i \sqrt{\gamma_{ex}} \hat{a}(t),
	\label{eq:InputOutputR}
\end{equation}
where $\hat{c}^{\mathrm{in}}(t)=\lim_{t_{i}\rightarrow -\infty} \int \frac{d\nu}{\sqrt{2\pi}} e^{-i \nu (t-t_{i})} \hat{c}_{\nu}(t_{i})$ and $\hat{c}^{\mathrm{out}}(t)=\lim_{t_{f}\rightarrow \infty} \int \frac{d\nu}{\sqrt{2\pi}} e^{i \nu (t_{f}-t)} \hat{c}_{\nu}(t_{f})$ are the noise operators at $t=-\infty$ (input) and $t=+\infty$ (output), respectively. The evolution of the cavity operators $\hat{a}(t)$ and $\hat{b}(t)$ is governed by the quantum Langevin equations
\begin{subequations}\label{eq:QLangevinE}
\begin{eqnarray}
	\partial_{t} \hat{a}(t)&=& (-i\nu_{a}-\frac{\gamma_{a}^{\prime}}{2}) \hat{a}(t)-i g \hat{b}(t)- i\sqrt{\gamma_{a}} \hat{a}^{\mathrm{in}}(t)-i \sqrt{\gamma_{ex}}\hat{c}^{\mathrm{in}}(t),~~  \label{eq:QLangevinE-a}\\
	\partial_{t} \hat{b}(t)&=& (-i\nu_{b}-\frac{\gamma_{b}}{2}) \hat{b}(t)-i g \hat{a}(t)- i \sqrt{\gamma_{b}} \hat{b}^{\mathrm{in}}(t),
	\label{eq:QLangevinE-b}
\end{eqnarray}
\end{subequations}
where $\gamma_{a}^{\prime}=\gamma_{a}+\gamma_{ex}$ and the definitions of $\hat{a}^{\mathrm{in}}(t)$ and $\hat{b}^{\mathrm{in}}(t)$ are similar to that of $\hat{c}^{\mathrm{in}}(t)$.  The input-output relation is found to be
\begin{align}
	\hat{c}^{\mathrm{out}}(\omega)=&\hat{c}^{\mathrm{in}}(\omega)-i\sqrt{\gamma_{ex} }G_{a}(\omega)\left[\sqrt{\gamma_{a}}\hat{a}^{\mathrm{in}}(\omega)+\sqrt{\gamma_{ex}} \hat{c}^{\mathrm{in}}(\omega)\right] \nonumber \\
	&-i\sqrt{\gamma_{ex} \gamma_{b}}G_{a}(\omega)g G^{(0)}_{b} \hat{b}^{\mathrm{in}}(\omega),
\label{eq:WaveguideO}
\end{align}
where  $G_{a}(\omega)=\frac{1}{\omega-\nu_{a}+i \frac{\gamma_{a}^{\prime}}{2}-g^{2}G_{b}^{(0)}(\omega)}$ is the dressed propagator of cavity $a$, and $G_{b}^{(0)}(\omega)=\frac{1}{\omega-\nu_{b}+i\frac{\gamma_{b}}{2}}$ is the free propagator of cavity $b$. The solution is in frequency domain after the Fourier transform $\hat{o}(\omega)=\int \frac{d t}{\sqrt{2\pi}} \hat{o}(t) e^{i \omega t}$. Comparing Eq. \eqref{eq:ScatteringTheory} with Eq. \eqref{eq:WaveguideO}, we find a correspondence between these two theories as  $\hat{o}_{1}=\hat{a}$,  $\hat{o}_{2}=\hat{b}$, $\gamma_{ex,j}=\gamma_{ex} \delta_{j,1}$, and $\hat{o}^{\mathrm{in}}_{1/2}(\nu)=-i\sqrt{2\pi \gamma_{a/b}}\hat{a}^{\mathrm{in}}(\nu)(\hat{b}^{\mathrm{in}}(\nu))$. For a given input state $\rho^{\mathrm{in}}$, Eq. \eqref{eq:WaveguideO} provides us a way to  calculate the output average of any waveguide operator $\hat{o}(\hat{c}_{\nu},\hat{c}^{\dagger}_{\nu})$ and hence the waveguide output state $\rho_{c}^{\mathrm{out}}$ (See Appendix \ref{sec:WignerR-app} for details).

Now to be specific we consider a Gaussian input state, which is the most commonly used in experiments. The output state must also be Gaussian since the scattering is a linear transform. The Gaussian state enables the exact calculation of the QFI. We assume that the free space and the waveguide are in the thermal equilibrium state with temperature $1/\beta$ (Boltzmann constant $k_{\mathrm{B}}$ taken as unity)  and the input signal is in the coherent state. The density matrix of the waveguide photons at frequency $\nu$ is $\rho_{c,\nu}^{\mathrm{in}}=\hat{D}(\alpha_{\nu})\rho_{c,\nu}^{T} \hat{D}^{\dagger}(\alpha_{\nu})$,
where $\rho_{c,\nu}^{T}=(1-e^{-\beta \nu})e^{-\beta \nu \hat{c}^{\dagger}_{\nu}\hat{c}_{\nu}}$ represents the thermal background photons in the waveguide, and $\hat{D}(\alpha_{\nu})=e^{\alpha_{\nu}\hat{c}^{\dagger}_{\nu}-\alpha_{\nu}^{\ast} \hat{c}_{\nu}}$ is the displacement operator which superimposes the coherent state on the  thermal background.  The free-space photons coupled to cavity $a$ and $b$ are in the thermal states $\rho^{T}_{a,\nu}$ and $\rho^{T}_{b,\nu}$, respectively. Thus the input state $\rho^{\mathrm{in}}=\bigotimes_{\nu}\rho_{a,\nu}^{T}\otimes \rho_{b,\nu}^{T}\otimes \rho_{c, \nu}^{\mathrm{in}}$. By using the input-output relation Eq. \eqref{eq:WaveguideO}, the waveguide output state $\rho^{\mathrm{out}}_{c}=\bigotimes_{\nu}\rho^{\mathrm{out}}_{c,\nu}$ is obtained (see Appendix \ref{sec:WignerR-app} for details). For the Gaussian output, the density matrix $\rho^{\mathrm{out}}_{c,\nu}$ and hence the QFI are fully determined by the expectation values and the second-order correlations of the quadrature operators $\hat{X}_{1,\nu}=\frac{1}{\sqrt{2}}(\hat{c}_{\nu}+\hat{c}^{\dagger}_{\nu})$ and $ \hat{X}_{2,\nu}= \frac{1}{i\sqrt{2}}(\hat{c}_{\nu}-\hat{c}^{\dagger}_{\nu})$. We denote the expectation values as $\bar{\mathbf{X}}_{\nu}=\left[ \langle \hat{X}_{1,\nu}\rangle, \langle \hat{X}_{2,\nu}\rangle \right]$ and the correlations as the covariance matrix $\left(\mathbb{C}_{\nu}\right)_{ij}=\frac{1}{2}\langle \hat{X}_{i,\nu}\hat{X}_{j,\nu}+\hat{X}_{j,\nu}\hat{X}_{i,\nu}\rangle -\langle \hat{X}_{i,\nu}\rangle \langle \hat{X}_{j,\nu}\rangle$. The results are
\begin{equation}
	\bar{\mathbf{X}}^{T}_{\nu}= \sqrt{2} \left[
	\begin{array}{c}
		\mathrm{Re}[\alpha_{\nu}(1-i\gamma_{ex} G_{a}(\nu))]\\
		\mathrm{Im}[\alpha_{\nu}(1-i \gamma_{ex} G_{a}(\nu))]
	\end{array}
	\right], ~
	\mathbb{C}_{\nu}= (\bar{n}_{\nu}+\frac{1}{2}) \mathbb{I},
\label{eq:AverageResults}
\end{equation}
where $\bar{n}_{\nu}=(e^{\beta \nu}-1)^{-1}$ is the average thermal
photon number. The identity form of $\mathbb{C}_{\nu}$ here is due to the particular coherent input state $\rho^{\mathrm{in}}_{c,\nu}$. It does not hold for general Gaussian input states. For example, off-diagonal elements exist for the squeezed input state. The QFI for the single-mode Gaussian state $\rho^{\mathrm{out}}_{c,\nu}$ reads (see Appendix \ref{sec:QFI-app} for details) \cite{Scutaru1998,Pinel2013},
\begin{equation}
	F^{\epsilon}_{\nu}=\frac{\mathrm{Tr}[(\mathbb{C}_{\nu}^{-1}\dot{\mathbb{C}}_{\nu})^{2}]}{2(1+P_{\nu}^{2})} + \frac{2 (\dot{P}_{\nu})^{2}}{1-P_{\nu}^{4}}+\dot{\bar{\mathbf{X}}}_{\nu}^{T}\mathbb{C}_{\nu}^{-1}\dot{\bar{\mathbf{X}}}_{\nu},
	\label{eq:FI-Gaussian}
\end{equation}
where the dot symbol denotes the derivative $\partial_{\epsilon}$ and  $P_{\nu}\equiv \mathrm{det}[2\mathbb{C}_{\nu}]^{-1/2}$ denotes the purity. The QFI for all the waveguide modes (which are taken as independent of each other) is $F^{\epsilon}=\int \frac{d\nu}{2 \pi} F^{\epsilon}_{\nu}$. Using Eq. \eqref{eq:AverageResults}, we obtain
\begin{equation}
	F^{\epsilon}= 4 \int \frac{d\nu}{2\pi}  \frac{|\alpha_{v}|^{2}}{2\bar{n}_{\nu}+1} \left| \frac{d S_{\nu}}{d \epsilon}\right|^{2},
	\label{eq:FI-CoupledCavity}
\end{equation}
where $S_{\nu}= \gamma_{ex} G_{a}(\nu)$ characterizes the scattering amplitude and the term $|\alpha_{\nu}|^{2}/(2 \bar{n}_{\nu}+1)$ characterizes the signal-to-noise ratio. The propagator has an explicit expression $G_{a}(\nu)=\frac{(\nu-\nu_{b}+i \frac{\gamma_{b}}{2})}{(\nu-\nu_{+})(\nu-\nu_{-})}$. Near the EP, each mode $\nu_{\pm}$ shows square root perturbation dependence, which makes the susceptibility divergent. However, the product $(\nu-\nu_{+})(\nu-\nu_{-})\approx(\nu-\frac{\epsilon+i(\bar{\gamma}+\frac{\gamma_{ex}}{2})}{2})^{2}-\gamma(g-\frac{|\gamma|}{2}-i \frac{\epsilon}{2})$ gives a smooth, linear perturbation dependence. Therefore the QFI $F^{\epsilon}$ shows no divergence at the EP.

\begin{figure}[tpb]
	\centering
	\includegraphics[width=1.0\columnwidth]{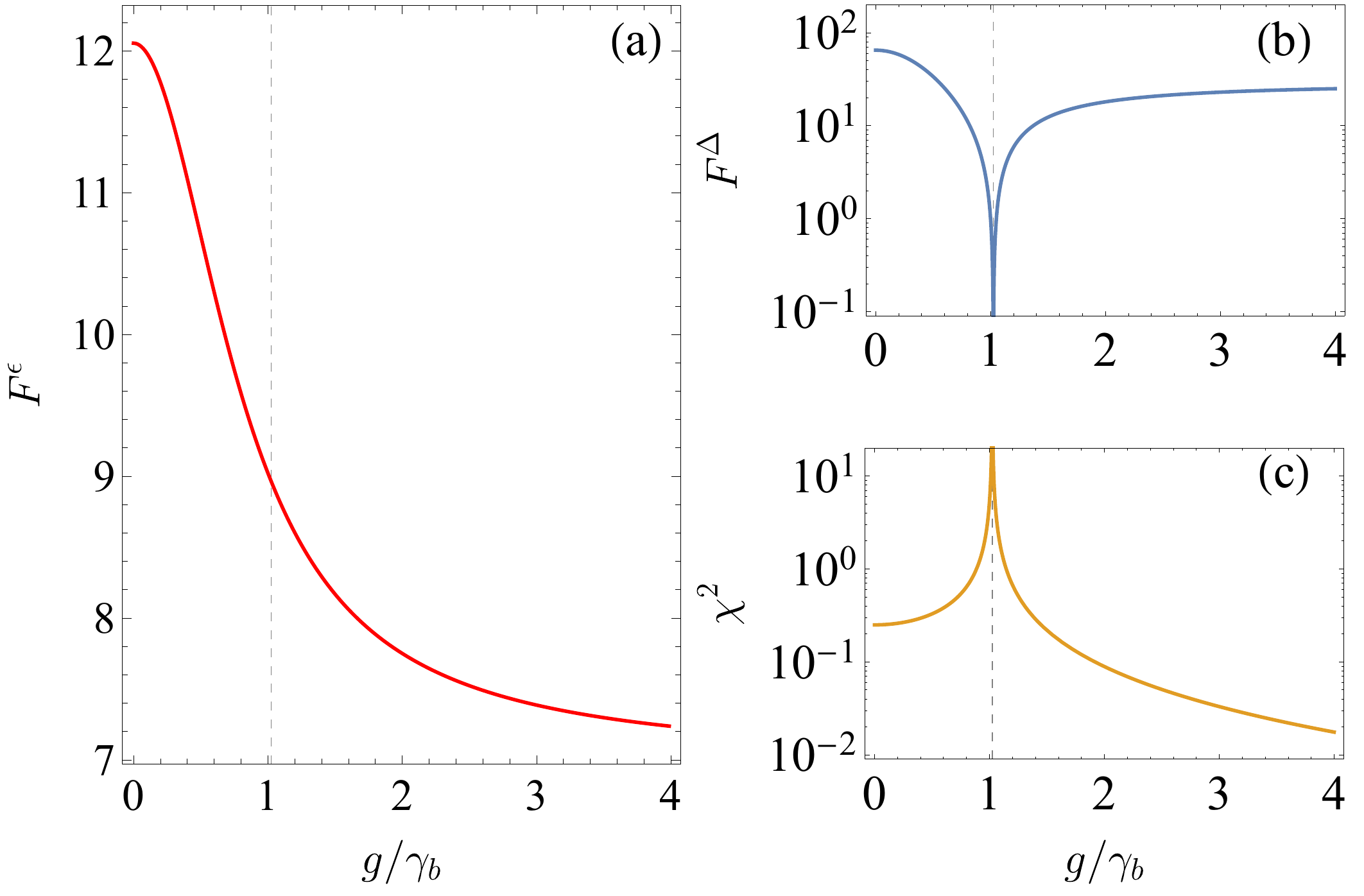}
	\caption{Numerical results for the coupled-cavity sensing near the EP.  (a) The QFI of cavity $a$ frequency $F^{\epsilon}$, (b) the QFI of the energy splitting $F^{\Delta}$, and (c) the susceptibility of the energy splitting $\chi^{2}$, as functions of the coupling strength $g$. The vertical dashed lines indicate the position of the EP.   Parameters are $\gamma_{b}=1.0$, $\gamma_{a}=5.0\gamma_{b}$, $\gamma_{ex}=0.1\gamma_{b}$, $\nu_{a}=\nu_{b}$, $\alpha=1000.0$, $\Gamma=200\gamma_{b}$, and $\beta\rightarrow\infty$.}
	\label{fig:FI-I}
\end{figure}

Figure \ref{fig:FI-I} presents the numerical results of the QFI and the energy splitting susceptibility as functions of the coupling strength $g$. Here, the input coherent state is assumed to have a spectrum $\alpha_{\nu}=\alpha \frac{\sqrt{2 \Gamma}}{\nu-\nu_{b}+i \frac{\Gamma}{2}}$, where $\alpha$ is the amplitude and the bandwidth $\Gamma \gg \gamma_{a},\gamma_{b}$. In the calculation, zero temperature is considered, i.e., $\bar{n}_{\nu}=0$, and the two cavities are tuned to resonance, i.e., $\nu_{a}=\nu_{b}$. The results reveal that $F^{\epsilon}$ is a smooth function of $g$ even at the EP (indicated by vertical dashed lines in Fig. \ref{fig:FI-I}).

To show that the absence of divergence of QFI at the EP is related to the state coalescence, we expand the QFI as
\begin{equation}
	F^{\epsilon}= 4\int \frac{d\nu}{2\pi} \frac{|\alpha_{\nu}|^{2}}{2\bar{n}_{\nu}+1}\left[ \left|\frac{\partial S_{\nu}}{\partial \Delta} \frac{d \Delta}{d \epsilon}\right|^{2}+2\mathfrak{R}\left(\frac{\partial S_{\nu}}{\partial \Delta} \frac{\partial S_{\nu}}{\partial \bar{\nu}}\frac{d \Delta}{d \epsilon} \frac{d \bar{\nu}}{d \epsilon} \right)+\left|\frac{\partial S_{\nu}}{\partial \bar{\nu}} \frac{d \bar{\nu}}{d \epsilon}\right|^{2}  \right],
\label{eq:QFI-expansion}
\end{equation}
from which, we define the QFI for the splitting $\Delta=(\nu_{+}-\nu_{-})$ as
\begin{equation}
	F^{\Delta}=4 \int \frac{d\nu}{2\pi} \frac{|\alpha_{\nu}|^{2}}{2\bar{n}_{\nu}+1}\left|\frac{\partial S_{\nu}}{\partial \Delta} \right|^{2}.
\label{eq:QFIofSpliting}
\end{equation}
It measures the available information in the output state $\rho_{\mathrm{c}}^{\mathrm{out}}$ to distinguish the energy splitting. From $\partial_{\Delta }S_{\nu}=\frac{(\nu+i \frac{\gamma_{b}}{2})\Delta}{(\nu-\nu_{+})^{2}(\nu-\nu_{-})^{2}}$, one can see that $F^{\Delta} \sim |\Delta|^{2}$ near the EP (see Fig. \ref{fig:FI-I} b). It reflects that the eigenstates are indistinguishable at the EP. Combining $F^{\Delta }$ with the divergent susceptibility $\chi^{2}$ (see Fig. \ref{fig:FI-I} c), we find that the susceptibility divergence is exactly counteracted by the vanishing QFI $F^{\Delta}$. Similar arguments apply to the second term in the expression of $F^{\epsilon}$ shown in Eq. \eqref{eq:QFI-expansion}. Thus the QFI $F^{\epsilon}$ is a smooth function around the EP.

\section{Active-passive cavity system}

By embedding a gain medium into cavity $b$, the decay rate $\gamma_{b}$ is effectively reduced and can even change the sign to realize an active cavity. Through this method, an effective active-passive coupled cavity system has been realized to study the $\mathcal{PT}$ symmetry \cite{Peng2014, Chang2014}.  It is interesting to know whether the EP in the active-passive system can enhance the sensitivity. The gain can be realized, e.g., by stimulated emission of a medium with population inversion. However, there exists a threshold that limits the maximal achievable gain rate. Above the threshold, the system will be in the lasing phase (a self-adaptive region) in which the effective decay rate description becomes invalid. In this study, we constrain the gain rate below the lasing threshold.

Below the threshold, the gain cavity works as an amplifier. The decay rate of the gain cavity due to the pumped gain medium becomes $\gamma_{b}^{\prime}=\gamma_{b}-4 S_{z} g^{2}_{G}/\kappa$ and the noise operator $\sqrt{\gamma^{\prime}_{b}} \hat{b}^{\mathrm{in}\prime}(\omega)=\sqrt{\gamma_{b}} \hat{b}^{\mathrm{in}}(\omega)+i \sqrt{\frac{4S_{z} g^2_{G}}{\kappa}} \hat{d}^{\mathrm{in} \dagger}(\omega)$, where $S_{z}\equiv N_{e}-N_{g}$ denotes the population inversion of the gain medium, $g_{G}$ is the cavity-gain medium coupling coefficient, $\kappa$ is the effective decay rate of the gain medium, and $\hat{d}^{\mathrm{in} \dagger}(\omega)$ is the noise operator induced by the gain medium with the average excitation number $\bar{n}_{d}=\frac{N_{g}}{S_Z}$ in the thermal state (see Appendix \ref{sec:APsystem-app} for details). The input-output theory, the waveguide output states, and the QFI for the passive-passive coupled cavity system are still valid here with only substitutions $\gamma_{b}\rightarrow \gamma_{b}^{\prime}$ and $\hat{b}^{\mathrm{in}}(\omega)\rightarrow \hat{b}^{\mathrm{in}\prime}(\omega)$. That yields
\begin{equation}
	\bar{\mathbf{X}}^{T}_{\nu}= \sqrt{2} \left[
	\begin{array}{c}
		\mathrm{Re}[\alpha_{\nu}(1-i\gamma_{ex} G_{a}(\nu))]\\
		\mathrm{Im}[\alpha_{\nu}(1-i \gamma_{ex} G_{a}(\nu))]
	\end{array}
	\right], ~
	\mathbb{C}_{\nu}= (\bar{n}^{\prime}_{\nu}+\frac{1}{2})\mathbb{I},
\label{eq:QuaVar-GP}
\end{equation}
where $\bar{n}^{\prime}_{\nu}=\bar{n}_{\nu}+\gamma_{ex} \frac{4S_{z}g_{G}^{2}}{\kappa}(\bar{n}_{\nu}+\frac{N_{e}}{S_{z}}) |g G_{a}(\nu) G_{b}^{(0)}(\nu)|^{2}$ is the average photon number modified by the gain medium. Then the QFI in Eq. \eqref{eq:FI-Gaussian} becomes
\begin{equation}
	F^{\epsilon}=\int \frac{d \nu}{2\pi} \left[4\frac{|\alpha_{\nu}|^{2}}{2\bar{n}^{\prime}_{\nu}+1} \left|\frac{d S_{\nu}}{d \epsilon}\right|^{2}+\frac{|\partial_{\epsilon}\bar{n}^{\prime}_{\nu}|^{2}}{\bar{n}^{\prime}_{\nu}(\bar{n}^{\prime}_{\nu}+1)}\right].
	\label{eq:QFI-GP}
\end{equation}
Below the threshold, both $G_{a}(\nu)=\frac{\nu-\nu_{b}+i \frac{\gamma^{\prime}_{b}}{2}}{(\nu-\nu_{+})(\nu-\nu_{-})}$ and $\bar{n}^{\prime}(\nu)=\bar{n}(\nu)+\frac{\gamma_{ex}g^{2} \frac{4 S_z g_{G}^{2}}{\kappa}}{|(\nu-\nu_{+})(\nu-\nu_{-})|^{2}}$ are well defined,  where $\nu_{\pm}=\bar{\nu}-i \frac{\gamma_{a}^{\prime}+\gamma_{b}^{\prime}}{4}\pm\sqrt{g^{2}+(\frac{\epsilon}{2}-i \frac{\gamma_{a}^{\prime}-\gamma_{b}^{\prime}}{4})^{2}}$ are the eigenvalues of the coefficient matrix of the active-passive coupled cavity system. Therefore, $F^{\epsilon}$ is a smooth function at the EP.

\begin{figure}[tpb]
	\centering
	\includegraphics[width=1.0\columnwidth]{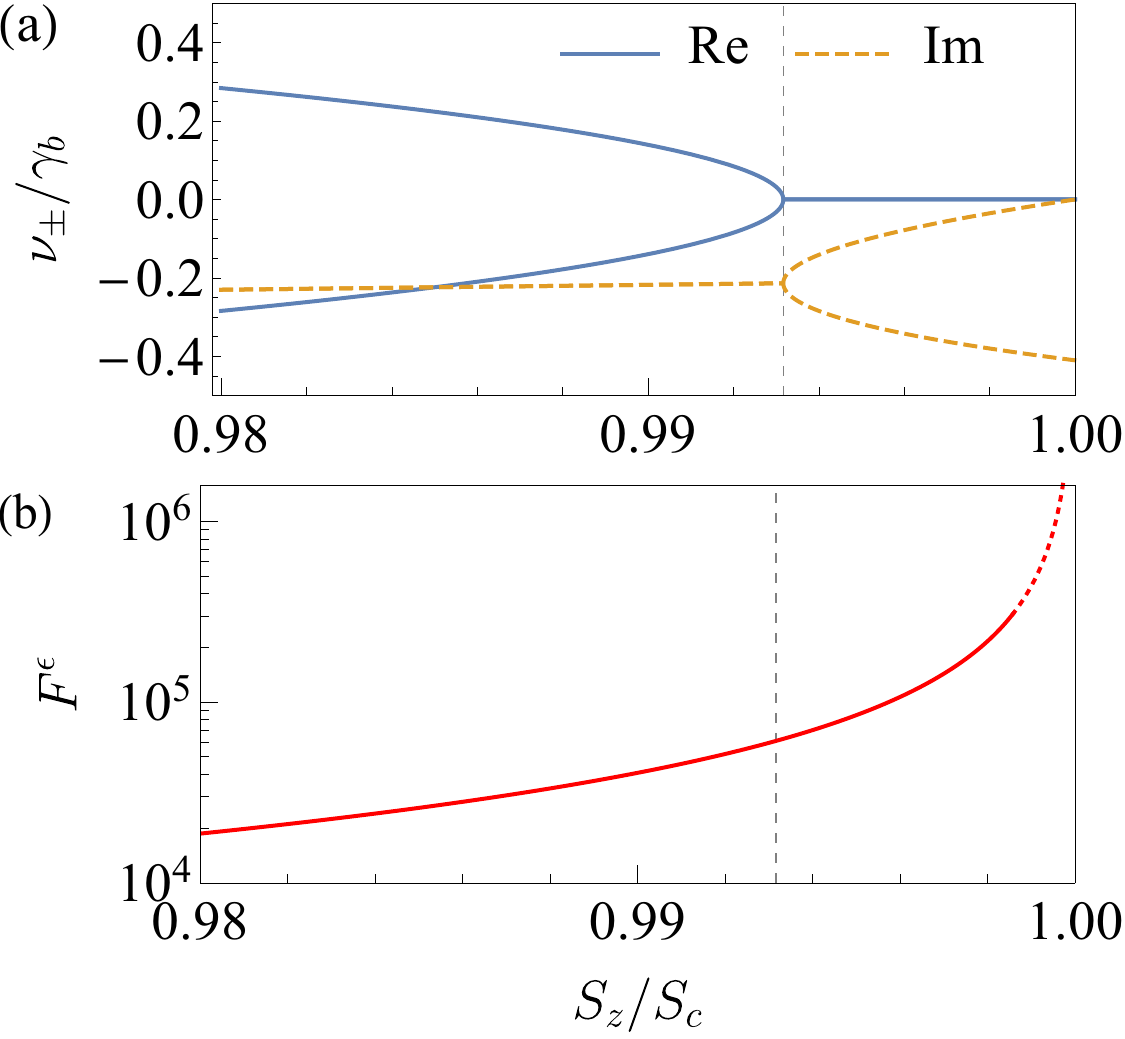}
	\caption{(a) The frequencies of the quasinomal modes and (b) the QFI as functions of $S_{z}/S_{c}$, where $S_{c}$ is the lasing threshold. The QFI is a smooth function of $S_{z}$ at the EP (indicated by the vertical dashed line). Near the threshold, the QFI diverges as revealed by the dotted line in (b). Results are all based on the linearized theory. Parameters are: $\gamma_{b}=1.0$, $\gamma_{a}=5.0\gamma_{b}$, $\gamma_{ex}=0.1\gamma_{b}$, $\nu_{a}=\nu_{b}$, $g=2.4\gamma_{b}$, $\kappa=100\gamma_{b}$, $N=2 \times 10^{12}$, $\gamma_{1}=0.01\gamma_{b}$, $g_{G}=10^{-5}\gamma_{b}$, $S_{c}=1.38\times 10^{12}$, and $\beta\rightarrow \infty$.}
	\label{fig:QFI-GP}
\end{figure}

Figure \ref{fig:QFI-GP} presents the numerical results of the quasinormal mode frequencies (a) and the QFI (b) as functions of $S_{z}$. The input state takes the same form as in Fig. \ref{fig:FI-I}. Figure \ref{fig:QFI-GP} (b) reveals that the QFI is a smooth function of the population inversion $S_{z}$ around the EP (indicated by the vertical dashed line). The enhancement of the QFI with increasing the population inversion is induced by the gain medium. The stronger the optical pumping is, the larger population inversion is induced, and the higher sensitivity is obtained.

In the linear theory, the QFI diverges at the lasing threshold, i.e., $S_{z}=S_{c}$. In Fig. \ref{fig:QFI-GP} (b), the divergent behavior is shown in the dotted line. However, the critical fluctuation neglected in linear description becomes important near the threshold, which may prevent the sensitivity from divergence. The discussion of the effects of the critical fluctuations is beyond the scope of this paper. Further increasing the pumping power, the coupled cavity system will exceed the threshold and enter the lasing phase. The EP, known as the $\mathcal{PT}$ phase transition point, occurs at the point  $g=\frac{\gamma^{\prime}_{a}}{2}$ and $\epsilon=0$ in the parametric space; when $2g>\gamma^{\prime}_{a}$, the system is in the $\mathcal{PT}$ symmetric lasing phase, where both modes are lasing; whereas when $2g<\gamma^{\prime}_{a}$, the $\mathcal{PT}$ symmetry breaks and the system is in the single mode lasing phase \cite{Feng2014, Hodaei2014, Hossein2016, Zhang2018a}. In contrast to the cases below the lasing threshold, a non-equilibrium phase transition occurs at the EP.  The conclusion revealed in Eq. \eqref{eq:FunctionalAnalysis}, that the enhancement of the QFI at the lasing transition is not caused by the divergence of the energy splitting susceptibility but the phase transition, can be generalized to this case. We can also understand this conclusion from the coalescence of the different quasinormal modes counteracts the susceptibility divergence at the EP.

\section{Conclusion and discussion}
We show that the exceptional point in a non-Hermitian sensing system does not dramatically enhance the sensitivity, since the coalescence of the different quasinormal modes counteracts the singular behavior of the mode splitting. This is verified in the passive-passive and active-passive coupled cavity systems through the exact calculation of the quantum Fisher information. This conclusion is valid for high-order EPs and other sensing schemes.

\textbf{Notes.} After completion of this work, we came across the paper [W. Langbein, arXiv:1801.05750], to whose conclusion ours is similar.

\begin{acknowledgments}
This work was supported by Hong Kong Research Grants Council and National Natural Science Foundation of China (Grant No. 11605094).
\end{acknowledgments}

\appendix
\section{Quantum scattering theory and QFI} \label{sec:QST-QFI-app}

The sensing process by the linear optical system can be described by a scattering process. We define
\begin{eqnarray}
	\hat{c}_{\nu}(t)&=&  \lim_{t^{\prime}\rightarrow -\infty} e^{i \hat{H} (t-t^{\prime})} e^{-i \hat{H}_{0} (t-t^{\prime})}\hat{c}^{\mathrm{in}}_{\nu}(t) e^{i \hat{H}_{0}(t-t^{\prime})} e^{-i \hat{H} (t-t^{\prime})},
	\label{eq:Evolution-app}
\end{eqnarray}
where $\hat{c}_{\nu}$ is the operator of the scattering photon,  $\hat{H}=\hat{H}_{0}+\hat{V}$ is the total Hamiltonian of the input photons and the sensing system with $\hat{H}_{0}$ being the free Hamiltonian and $\hat{V}$ being the interaction Hamiltonian between input photons and the sensing system. Taking a derivative of both sides of Eq. \eqref{eq:Evolution-app} with respect to time $t$, we get
\begin{equation}
	\partial_{t}\hat{c}_{\nu}(t)= -i \nu \hat{c}_{\nu} (t)+i \lim_{t^{\prime}\rightarrow -\infty}  e^{i \hat{H} (t-t^{\prime})} [\hat{V}, \hat{c}^{\mathrm{in}}_{\nu}(t^{\prime})] e^{-i \hat{H} (t-t^{\prime})},
	\label{eq:Derivative-app}
\end{equation}
where $\partial_{t} \hat{c}^{\mathrm{in}}_{\nu}(t)=-i \nu \hat{c}^{\mathrm{in}}_{\nu}(t)$ is used. A formal time integration of Eq. \eqref{eq:Derivative-app} yields
\begin{equation}
	\hat{c}_{\nu}(t)= \hat{c}^{\mathrm{in}}_{\nu}(t)+i \int^{\infty}_{0} d\tau  e^{i  \hat{H} \tau} [\hat{V}, \hat{c}^{\mathrm{in}}_{\nu}(t-\tau)]  e^{-i \hat{H} \tau}.
	\label{eq:InputT-app}
\end{equation}
Utilizing the formulas that $e^{i \hat{H}t}\hat{o} e^{-i \hat{H}t}=e^{i t\hat{H}\wedge} \hat{o}$ and $\hat{c}^{\mathrm{in}}_{v}(t-\tau)=\hat{c}_{\nu} e^{-i \nu (t-\tau)}$, we simplify the result to
\begin{equation}
	\hat{c}_{\nu}(t)= \hat{c}^{\mathrm{in}}_{\nu}(t)- e^{-i \nu t} \frac{1}{\nu+\hat{H}\wedge} [\hat{V},\hat{c}_{\nu}].
	\label{eq:Simplify-app}
\end{equation}
With the formula $\hat{c}^{\mathrm{out}}_{\nu}(t)= \hat{\Omega}^{\dagger}_{-} \hat{c}_{\nu}(t) \hat{\Omega}_{-}$, the input-output relation is obtained
\begin{align}
	\hat{c}^{\mathrm{out}}_{\nu}(t)=& \hat{c}^{\mathrm{in}}_{\nu}(t)-\lim_{t^{\prime}\rightarrow \infty} e^{-i \nu t^{\prime}} e^{i \hat{H}_{0}(t-t^{\prime})} \left(\frac{1}{\nu+\hat{H}\wedge} [\hat{V},\hat{c}_{\nu}]\right) e^{i \hat{H}_{0}(t-t^{\prime})}, \nonumber \\
	=& \hat{c}^{\mathrm{in}}_{\nu}(t)-e^{-i\nu t}\left(\frac{1}{\nu+\hat{H}\wedge} [\hat{V},\hat{c}_{\nu}]\right)^{\mathrm{in}}_{\nu},
	\label{eq:InputOutput-St-app}
\end{align}
where $\left(\hat{o} \right)^{\mathrm{in}}_{\nu}=\int \frac{d t}{\sqrt{2\pi}} e^{i \nu t} \hat{o}^{\mathrm{in}}(t) $ denotes frequency $\nu$ component of contribution. We consider a general case of linear systems. The Hamiltonian reads
\begin{eqnarray}
	\hat{H}_{0}&=& \int d\nu \nu \hat{c}^{\dagger}_{\nu} \hat{c}_{\nu}+\sum_{lk}\hat{o}^{\dagger}_{l}M_{lk}\hat{o}_{k},\\
	\hat{V}&=& \int \frac{d \nu}{\sqrt{2\pi}}\sum_{j} \sqrt{\gamma_{ex,j}} (\hat{o}^{\dagger}_{j} \hat{c}_{\nu}+h.c.),
	\label{eq:ST-Operator-app}
\end{eqnarray}
where $\hat{o}_{l}$ are linear operators of the sensing system satisfying the  commutator $[\hat{o}^{\dagger}_{l}, \hat{o}_{k}]=\delta_{l,k}$ and $\gamma_{ex,j}$ is the coupling strength between input photons and the $j$-th mode of the sensing system. Taking the interaction term as perturbation and expanding it to the second order, we get
\begin{align}
	\hat{c}^{\mathrm{out}}_{\nu}\approx& \hat{c}^{\mathrm{in}}_{\nu}+\sum_{lj}(M_{\nu}^{-1})_{lj}\left[ \sqrt{\frac{\gamma_{ex,j}}{2\pi}}\hat{o}^{\mathrm{in}}_{l}(\nu)+\int \frac{d\nu^{\prime}}{2\pi} \frac{\sqrt{\gamma_{ex,l}\gamma_{ex,j}}\hat{c}^{\mathrm{in}}_{\nu^{\prime}}}{\nu-\nu^{\prime}+i\delta^{+}}\right], \nonumber \\
	\approx & \hat{c}^{\mathrm{in}}_{\nu}+\sum_{lj}(M_{\nu}^{-1})_{lj}\sqrt{\gamma_{ex,j}} [\frac{\hat{o}^{\mathrm{in}}_{l} (\nu)}{\sqrt{2\pi}}-i\sqrt{\gamma_{ex,l}}\hat{c}^{\mathrm{in}}_{\nu}].
\label{eq:ScatteringTheory-app}
\end{align}

The output state takes the form
\begin{equation}
	\rho(\epsilon)=\bigotimes_{\nu} P^{\nu}_{n m} \frac{(\hat{c}^{\mathrm{out} \dagger}_{\nu})^{n}}{\sqrt{n!}}|0\rangle \langle 0|\frac{(\hat{c}^{\mathrm{out}}_{\nu})^{m}}{\sqrt{m!}},
	\label{eq:OutputState-app}
\end{equation}
where $P^{\nu}_{nm}=\langle n_{\nu}|\rho^{\mathrm{in}}|m_{\nu}\rangle$ is the density matrix element of the input state. Here we suppose the input state is a product state of different frequency modes. A small disturbance $\delta \epsilon$ of the sensing system changes the output state to
\begin{equation}
	\rho(\epsilon+\delta\epsilon)=\rho(\epsilon)+ \int \frac{d \nu}{\sqrt{2\pi}} \sum_{lj}\frac{\partial \rho(\epsilon)}{\partial (M^{-1}_{\nu})_{lj}}\frac{d (M^{-1}_{\nu})_{lj}}{d \epsilon} \delta \epsilon +\hat{O}(\delta \epsilon^2).
\end{equation}
Applying this formula into the definition of Bures distance defined in Eq. \eqref{eq:BuresDistance} yields
\begin{equation}
	d_{B}^{2}[\rho(\epsilon),\rho(\epsilon+\delta \epsilon)]=2-2\mathrm{Tr}\sqrt{\rho^2(0)+\rho^{1/2}(\epsilon) \delta \rho (\epsilon)\rho^{1/2}(\epsilon)},
	\label{eq:BuresDistance-app}
\end{equation}
where $\delta \rho(\epsilon)=\rho(\epsilon+\delta \epsilon)-\rho(\epsilon)$. Substituting Eq. \eqref{eq:BuresDistance-app} into the definition of QFI and representing the density matrix $\rho(\epsilon)$ in its eigenbasis, we get the expression of QFI as
\begin{align}
	F^{\epsilon}=& \lim_{\delta\epsilon\rightarrow 0}\frac{8}{\delta\epsilon^{2}} \left[ 1-\sum_{i} p_{\alpha} \sqrt{1+\frac{\epsilon}{p_{\alpha}} \langle u_{\alpha}|\partial_{\epsilon} \rho(\epsilon)|u_{\alpha}\rangle+\epsilon^{2}\sum_{\beta\neq \alpha} \frac{p_{\beta}|\langle u_{\alpha}|\partial_{\epsilon} \rho(\epsilon)|u_{\beta}\rangle|^{2}}{p_{\alpha}(p_{\alpha}^{2}-p_{\beta}^{2})}}\right]\nonumber \\
	=&2\sum_{\alpha,\beta} \frac{|\langle \mu_{\alpha} |\partial_{\epsilon}\rho(\epsilon)|\mu_{\beta}\rangle|^2}{p_{_\alpha}+p_{\beta}},
\label{eq:QFI-app}
\end{align}
where $|\mu_{\alpha}\rangle$ is the $\alpha$-th eigenstate of $\rho(\epsilon)$ with the population $p_{\alpha}$.

\section{Wigner representation of the output state} \label{sec:WignerR-app}

The input-output formula in Eq. \eqref{eq:WaveguideO} provides us a way to calculate the output average of an arbitrary waveguide operator $\hat{o}$. The key steps are as follows: first, make a decomposition of $\hat{o}$ in terms of $\hat{c}_{\nu} $ and $\hat{c}^{\dagger}_{\nu}$ as $\hat{o}(\hat{c}_{\nu},\hat{c}^{\dagger}_{\nu})$; then, transform the output average to the input average through the Schr\"{o}dinger-Heisenberg picture transformation
\begin{equation}
	\langle \hat{o}\rangle_{\mathrm{out}}=\langle \hat{o}(\hat{c}^{\mathrm{out}}(\omega)e^{-i\omega t}, \hat{c}^{\mathrm{out} \dagger}(\omega)e^{i\omega t})\rangle_{\mathrm{in}};
\label{eq:OperatorAverage}
\end{equation}
finally, substitute the output operators in terms of the input operators by using Eq. \eqref{eq:WaveguideO} and make the input average. The output state $\rho^{\mathrm{out}}_{c}=\bigotimes_{\nu}\rho^{\mathrm{out}}_{c,\nu}$ is constructed from the Wigner-Weyl representation
\begin{equation}
	\rho^{\mathrm{out}}_{c,\nu}=\pi \int dz^{2}  W_{\nu}(z,z^{\ast}) \Xi(z-\hat{c}_{\nu}, z^{\ast}-\hat{c}_{\nu}^{\dagger}),
	\label{eq:WingerRep}
\end{equation}
where $\Xi(z-\hat{c}_{\nu}, z^{\ast}-\hat{c}^{\dagger}_{\nu})=\int \frac{d\xi^{2}}{\pi^{2}} e^{\xi^{\ast}(z-\hat{c}_{\nu})-h.c.}$ is the characteristic function and $W_{\nu}(z, z^{\ast})=\langle\Xi^{\dagger}(z-\hat{c}_{\nu}, z^{\ast}-\hat{c}_{\nu}^{\dagger})\rangle_{\mathrm{out}}$ is the Wigner function.

\section{QFI for the single mode Gaussian state} \label{sec:QFI-app}

The Wigner function of a Gaussian state has the form
\begin{equation} W_{\nu}(z,z^{\ast})=\frac{e^{-\frac{1}{2}(\mathbf{X}_{\nu}-\bar{\mathbf{X}}_{\nu})^{T}\mathbb{C}^{-1}_{\nu}(\mathbf{X}_{\nu}-\bar{\mathbf{X}}_{\nu})}}{\pi|\mathrm{det}[\mathbb{C}_{\nu}]|^{1/2}},
\label{eq:WignerFunction-app}
\end{equation}
where $\mathbf{X}_{\nu}= [\frac{z+z^{\ast}}{\sqrt{2}}, \frac{z-z^{\ast}}{i \sqrt{2}}]$ is the quadrature with the mean value $\bar{\mathbf{X}}_{\nu}= [\langle \hat{X}_{1,\nu}\rangle, \langle \hat{X}_{2,\nu} \rangle]$ and the covariance matrix $\left(\mathbb{C}_{\nu}\right)_{ij}=\frac{1}{2}\langle \hat{X}_{i,\nu} \hat{X}_{j,\nu}+\hat{X}_{j,\nu}\hat{X}_{i,\nu}\rangle -\langle \hat{X}_{i,\nu}\rangle \langle \hat{X}_{j,\nu}\rangle$. Here the quadrature operators are defined as $\hat{X}_{1,\nu}=\frac{\hat{c}_{\nu}+\hat{c}^{\dagger}_{\nu}}{\sqrt{2}}$ and $ \hat{X}_{2,\nu}= \frac{\hat{c}_{\nu}-\hat{c}^{\dagger}_{\nu}}{i\sqrt{2}}$, and  $\langle \cdot \rangle$ denotes the average over the output state.  For any two single-mode Gaussian states $\rho^{\mathrm{out}}_{c,\nu}(\epsilon)$ and $\rho^{\mathrm{out}}_{c,\nu}(\epsilon+\delta \epsilon)$ with mean values $\bar{\mathbf{X}}_{\nu}(\epsilon)$ and $\bar{\mathbf{X}}_{\nu}(\epsilon+\delta \epsilon)$ and covariance matrices $\mathbb{C}_{\nu}(\epsilon)$ and $\mathbb{C}_{\nu}(\epsilon+\delta \epsilon)$, Ref. \cite{Scutaru1998} gives an exact formula of the fidelity as
\begin{equation}
	\mathcal{F}(\rho^{\mathrm{out}}_{c,\nu}(\epsilon), \rho^{\mathrm{out}}_{c,\nu}(\epsilon+\delta \epsilon))=\frac{2 \exp\left(-\frac{1}{4} \delta \bar{\mathbf{X}}_{\nu}^{T} (\bar{\mathbb{C}}_{\nu})^{-1} \delta \bar{\mathbf{X}}_{\nu}\right)}{\sqrt{\det[4\mathbb{C}_{\nu}(\epsilon)]+A}-\sqrt{A}},
\label{eq:QFI-app}
\end{equation}
where $A=(\det[2\mathbb{C}_{\nu}(\epsilon+\delta\epsilon)]-1)(\det[2\mathbb{C}_{\nu}(\epsilon)]-1)$, $\delta \bar{\mathbf{X}}_{\nu}=\bar{\mathbf{X}}_{\nu}(\epsilon+\delta \epsilon)-\bar{\mathbf{X}}_{\nu}(\epsilon)$, and $\bar{\mathbb{C}}_{\nu}=\frac{1}{2}[\mathbb{C}_{\nu}(\epsilon)+\mathbb{C}_{\nu}(\epsilon+\delta\epsilon)]$. Based on this result, Ref. \cite{Pinel2013} presents a concise result of the QFI, namely
\begin{equation} F^{\epsilon}_{\nu}=\frac{\mathrm{Tr}[(\mathbb{C}_{\nu}^{-1}\dot{\mathbb{C}}_{\nu})^{2}]}
{2(1+P_{\nu}^{2})} + \frac{2 (\dot{P}_{\nu})^{2}}{1-P_{\nu}^{4}}+\dot{\bar{\mathbf{X}}}_{\nu}^{T}\mathbb{C}_{\nu}^{-1}\dot{\bar{\mathbf{X}}}_{\nu},
	\label{eq:FI-Gaussian-app}
\end{equation}
where $\bar{\mathbf{X}}_{\nu}=\bar{\mathbf{X}}_{\nu}(\epsilon)$, $\mathbb{C}_{\nu}=\mathbb{C}_{\nu}(\epsilon)$, the symbol dot denotes the derivative $\partial_{\epsilon}$, and $P_{\nu}\equiv \mathrm{det}[2\mathbb{C}_{\nu}]^{-1/2}$ denotes the purity.

\section{Active-passive coupled cavity system} \label{sec:APsystem-app}
Theoretically, the gain medium is treated as an ensemble of two level systems (TLSs) $\sum_{l=1}^{N} \hat{\sigma}_{l}^{j}$, where $\hat{\sigma}_{l}^{j}$ with $j=(x,y,z)$ are the Pauli matrices of the $l$-th TLS and $N$ is the total number of TLSs in the ensemble.  With the inhomogeneous broadening taken into consideration, the frequency splitting reads $\sum_{l}(\nu_{b}+\delta\nu_{l}) \hat{\sigma}_{l}^{z}$, where $\delta\nu_{l}$ denotes the stochastic fluctuation of the splitting. The central splitting is tuned resonant with the cavity frequency $\nu_{b}$. Due to the interaction with the free space electromagnetic field and the optical pumping field, each TLS relaxes from the upper to the lower level with rate $\gamma_{1}$ and is pumped from the lower to the upper level with rate $\mathit{w}_{p}$. The TLSs homogeneously couple to cavity $b$ through the dipole interaction $\sum_{l} g_{G} (\hat{\sigma}_{l}^{+}\hat{b}+\hat{b}^{\dagger}\hat{\sigma}_{l}^{-})$. Here, the rotating wave approximation is adopted. The Langevin equations for the TLS ensemble are
\begin{eqnarray}
	\partial_{t}\hat{S}_{-}(t)&=& -(i\nu_{b}+\frac{\kappa}{2})\hat{S}_{-}(t)+i g_{G}\hat{S}_{z}(t) \hat{b}(t) -i \hat{\xi}_{-}(t), \\
	\partial_{t}\hat{S}_{z}(t)&=& -(\mathit{w}_{p}+\gamma_{1})\hat{S}_{z}(t)-i2g_{G} [\hat{S}_{+}(t) \hat{b}(t)-\hat{b}^{\dagger}(t)\hat{S}_{-}(t)]\nonumber \\
	&&-N(\mathit{w}_{p}-\gamma_{1})-i \hat{\xi}_{z} (t),
	\label{eq:LangevianTLE}
\end{eqnarray}
where $\hat{S}_{\alpha}(t)=\sum_{l} \hat{\sigma}_{l}^{\alpha}(t)$ (with $\alpha=z$ or $-$) are the operators of the gain medium with noise terms $\hat{\xi}_{\alpha}(t)$, and $\kappa=1/T^{\ast}_{2}+\gamma_{1}+\mathit{w}_{p}$ denotes the decay rate of the $\hat{S}_{-}$ with $1/T_{2}^{\ast}$ being the inhomogeneous broadening.

Below the lasing threshold, the operator $\hat{S}_{z}$ is well approximated by its mean-field average, i.e., $\hat{S}_{z}(t)\approx S_{z}\equiv N_{e}-N_{g}= N\frac{\mathit{w}_{p}-\gamma_{1}}{\mathit{w}_{p}+\gamma_{1}}-2i g_{G}\langle \hat{S}_{+}\hat{b} -\hat{b}^{\dagger} \hat{S}_{-} \rangle$, where $N_{e}$ and $N_{g}$ denote the numbers of TLSs in the upper and lower levels, respectively, and $\langle \cdot\rangle$ denotes the average over the steady state. Applying the Holstein-Primakoff transformation, one can get an effective bosonic description of the ensemble as $\hat{S}_{-}(t)\approx \sqrt{|S_{z}|} \hat{d}^{\dagger}(t)$ or $\sqrt{|S_{z}|}\hat{d}(t)$ corresponding to $S_{z}>0$ or $S_{z}<0$, respectively. In this paper, we focus on the case $S_{z}>0$. The Langevin equations for cavity fields and the spin ensemble are
\begin{eqnarray}
	\partial_{t}\hat{a}(t)&=& -(i\nu_{a}+\frac{\gamma_{a}^{\prime}}{2})\hat{a}(t)-ig \hat{b}(t)-i \sqrt{\gamma_{a}} \hat{a}^{\mathrm{in}}(t)-i \sqrt{\gamma_{ex}}\hat{c}^{\mathrm{in}}(t), \nonumber \\
	\partial_{t}\hat{b}(t)&=& -(i\nu_{b}+\frac{\gamma_{b}}{2})\hat{b}(t)-ig \hat{a}(t)-i \sqrt{S_{z}} g_{G} \hat{d}^{\dagger}(t)-i \sqrt{\gamma_{b}} \hat{b}^{\mathrm{in}}(t), \nonumber\\
	\partial_{t} \hat{d}^{\dagger}(t)&=& -(i\nu_{b}+\frac{\kappa}{2})\hat{d}^{\dagger}(t)+i \sqrt{S_{z}} g_{G}\hat{b}(t) +i \sqrt{\kappa} \hat{d}^{\mathrm{in} \dagger}(t),
	\label{eq:Langevin-AP-app}
\end{eqnarray}
where the definitions $\gamma_{a}^{\prime}=\gamma_{a}+\gamma_{ex}$, $\hat{a}^{\mathrm{in}}(t)$, $\hat{b}^{\mathrm{in}}(t)$, and $\hat{c}^{\mathrm{in}}(t)$ are similar to Eq. \eqref{eq:QLangevinE} and $\hat{d}^{\mathrm{in} \dagger}(t)\equiv-\frac{\hat{\xi}_{-}(t)}{\sqrt{S_{z} \kappa}}\approx \sqrt{N_{e}\over S_{z}}\hat{d}^{\mathrm{in} \dagger}_{e}(t)-\sqrt{N_{g}\over S_{z}}\hat{d}_{g}^{\mathrm{in}}(t)$ is the linearized noise operator of the gain medium. The population inversion $S_{z}$ is obtained by solving the self-consistent equation
\begin{equation}
	S_{z}=N\frac{\mathit{w}_{p}-\gamma_{1}}{\mathit{w}_{p}+\gamma_{1}} - i \frac{2\sqrt{S_{z}} g_{G}}{w_{p}+\gamma_{1}} \lim_{t\rightarrow \infty}\langle (\hat{d}(t)\hat{b}(t)-h.c.)\rangle_{\mathrm{in}},
	\label{eq:SelfConsistentEq}
\end{equation}
where $\langle \cdot \rangle_{\mathrm{in}}$ stands for the average over the input state. It should be noticed that both $\hat{b}(t)$ and $\hat{c}(t)$ are $S_{z}$ dependent.

By the Fourier transform $\hat{o}(\omega)=\int \frac{dt}{ \sqrt{2\pi}}e^{i\omega t}\hat{o}(t)$, Eq. \eqref{eq:Langevin-AP-app} becomes
\begin{subequations}
\begin{eqnarray}
	\hat{a}(\omega)&=&  G_{a}(\omega)\left[\sqrt{\gamma_{a}}\hat{a}^{\mathrm{in}}(\omega)+\sqrt{\gamma_{ex}}\hat{c}^{\mathrm{in}}(\omega)+g G^{(1d)}_{b}(\omega) \sqrt{\gamma_{b}} \hat{b}^{\mathrm{in}} (\omega)\right. \nonumber \\
	&& \left. - g G^{(1d)}_{b}(\omega) \sqrt{S_{z}} g_{G} G_{d}^{(0)} (\omega)\sqrt{\kappa} \hat{d}^{\mathrm{in} \dagger}(\omega)\right], \label{eq:Spectrum-a-app} \\
	\hat{b}(\omega)&=& G_{b}(\omega)\left[ \sqrt{\gamma_{b}} \hat{b}^{\mathrm{in}}(\omega)+g G^{(0)}_{a}(\omega) (\sqrt{\gamma_{a}}\hat{a}^{\mathrm{in}}(\omega)+\sqrt{\gamma_{ex}}\hat{c}^{\mathrm{in}}(\omega))\right. \nonumber \\
	&& \left. -   \sqrt{S_{z}} g_{G} G^{(0)}_{d}(\omega) \sqrt{\kappa} \hat{d}^{\mathrm{in} \dagger}(\omega)\right], \label{eq:Spectrum-b-app}\\
	\hat{d}^{\dagger}(\omega)&=& G_{d}(\omega)\left [-\sqrt{\kappa}  \hat{d}^{\mathrm{in} \dagger}(\omega) -   g G^{(1a)}_{b}(\omega)\sqrt{\gamma_{b}}\hat{b}^{\mathrm{in}}(\omega)\right. \nonumber \\
	&&\left. -\sqrt{S_{z}} g_{G} G^{(1a)}_{b}(\omega) g G_{a}^{(0)}(\omega) (\sqrt{\gamma_{a}}\hat{a}^{\mathrm{in}}(\omega)+\sqrt{\gamma_{ex}}\hat{c}^{\mathrm{in}}(\omega))\right], \label{eq:Spectrum-s-app}
\end{eqnarray}
\end{subequations}
where the free propagators of cavity $a$, cavity $b$, and the gain-medium are in turn  $G_{a}^{(0)}(\omega)=\frac{1}{\omega-\nu_{a}+i \frac{\gamma_{a}^{\prime}}{2}}$, $G_{b}^{(0)}(\omega)=\frac{1}{\omega-\nu_{b}+i \frac{\gamma_{b}}{2}}$, and $G^{(0)}_{d}(\omega)=\frac{1}{\omega-i\nu_{b}+i \frac{\kappa}{2}}$;  the dressed propagators read $G_{b}^{(1a)}(\omega)=\frac{1}{\omega-\nu_{a}+i\frac{\gamma_{b}}{2}-g^{2} G_{a}^{(0)}(\omega)}$, $G_{b}^{(1d)}(\omega)=\frac{1}{\omega-\nu_{b}+i \frac{\gamma_{b}}{2}+S_{z} g_{G}^{2}G_{d}^{(0)}}$, $G_{a}(\omega)=\frac{1}{\omega-\nu_{a}+i \frac{\gamma_{a}^{\prime}}{2}-g^{2}G_{b}^{(1d)}(\omega)}$, $G_{b}(\omega)=\frac{1}{\omega-\nu_{b}+i \frac{\gamma_{b}}{2}-g^2 G_{a}^{(0)}(\omega)+{S_{z}} g_{G}^{2}G_{d}^{(0)}(\omega)}$, and $G_{d}(\omega)=\frac{1}{\omega-\nu_{b}+i \frac{\kappa}{2}+{S_{z}} g_{G}^{2}G_{b}^{(1a)}(\omega)}$. The waveguide output operator $\hat{c}^{\mathrm{out}}(\omega)$ follows the input-output relationship
\begin{equation}
	\hat{c}^{\mathrm{out}}(\omega)=\hat{c}^{\mathrm{in}}(\omega)-i \sqrt{\gamma_{ex}} \hat{a}(\omega).
	\label{eq:InputOutputR-app}
\end{equation}
Combining Eq. \eqref{eq:InputOutputR-app} with Eq. \eqref{eq:Spectrum-a-app}, we get
\begin{eqnarray}
	\hat{c}^{\mathrm{out}}(\omega)&=& \hat{c}^{\mathrm{in}}(\omega)-i \sqrt{\gamma_{ex}} G_{a}(\omega)(\sqrt{\gamma_{a}}\hat{a}^{\mathrm{in}}(\omega)+\sqrt{\gamma_{ex}}\hat{c}^{\mathrm{in}}(\omega)) \nonumber \\
	& &-i\sqrt{\gamma_{ex} \gamma_{b}} G_{a}(\omega) g G^{(1d)}_{b}(\omega)  \hat{b}^{\mathrm{in}} (\omega) \nonumber \\
	&& -i\sqrt{\gamma_{ex}\kappa} G_{a}(\omega)g G^{(1d)}_{b}(\omega) \sqrt{S_{z}} g_{G} G_{c}^{(0)} (\omega) \hat{d}^{\mathrm{in} \dagger}(\omega). ~~
	\label{eq:Output-app}
\end{eqnarray}
The construction of the waveguide output state $\rho^{\mathrm{out}}_{c}=\bigotimes_{\nu} \rho^{\mathrm{out}}_{c,\nu}$ follows the same procedures in the passive-passive  case (see Appendix \ref{sec:WignerR-app} for details).

Considering the same input state as in the passive-passive case, we have $\rho^{\mathrm{in}}=\bigotimes_{\nu} \rho^{\mathrm{in}}_{c,\nu}\otimes \rho^{T}_{a,\nu}\otimes \rho^{T}_{b,\nu}\otimes \rho^{\mathrm{in}}_{d,\nu}$ with $\rho^{\mathrm{in}}_{c,\nu}=\hat{D}(\alpha_{\nu}) \rho^{T}_{c,\nu} \hat{D}^{\dagger}(\alpha_{\nu})$; the $\nu$-frequency noise field associated with the gain medium is in the vacuum state $\rho^{\mathrm{in}}_{d,\nu}=|0_{e},0_{g}\rangle \langle 0_{e}, 0_{g}|$ for the operator $\hat{d}^{\dagger}_{\nu}$ so that the average excitation number $\bar{n}_{d,\nu}=\langle \hat{d}^{\dagger}_{\nu}\hat{d}_{\nu}\rangle_{\mathrm{in}}=\frac{N_{g}}{S_{z}}$. The waveguide output state $\rho^{\mathrm{out}}_{c}=\bigotimes_{\nu}\rho^{\mathrm{out}}_{c,\nu}$ takes a Gaussian form. As noted in Appendix \ref{sec:QFI-app}, the Gaussian state $\rho^{\mathrm{out}}_{c,\nu}$ is fully determined by the expectation values $\bar{\mathbf{X}}_{\nu}=[ \langle \hat{X}_{1,\nu}\rangle, \langle \hat{X}_{2,\nu}\rangle]$ and the covariance matrix $\left(\mathbb{C}_{\nu}\right)_{ij}=\frac{1}{2}\langle \left(\hat{X}_{i,\nu} \hat{X}_{j,\nu}+\hat{X}_{j,\nu}\hat{X}_{i,\nu}\right)\rangle -\langle \hat{X}_{i,\nu}\rangle \langle \hat{X}_{j,\nu}\rangle$. For the input state $\rho^{\mathrm{in}}$, we have
\begin{subequations}\label{eq:AverageResults-app}
\begin{eqnarray}
	\bar{\mathbf{X}}^{T}_{\nu}&=& \sqrt{2} \left[
	\begin{array}{c}
		\mathrm{Re}[\alpha_{\nu}(1-i\gamma_{ex} G_{a}(\nu))]\\
		\mathrm{Im}[\alpha_{\nu}(1-i \gamma_{ex} G_{a}(\nu))]
	\end{array}
\right], ~ ~\Sigma_{\nu}= (\bar{n}^{\prime}_{\nu}+\frac{1}{2})\mathbb{I},
\end{eqnarray}
\end{subequations}
where $\bar{n}^{\prime}_{\nu}=\bar{n}_{\nu}+(\bar{n}_{\nu}+\frac{N_{e}}{S_{z}})\gamma_{ex} \kappa|G_{a}(\nu)g G_{b}^{(1d)}(\nu)\sqrt{S_{z}} g_{G} G_{d}^{(0)}(\nu)|^{2}$ is the gain medium modified average photon number.

The QFI of the parameter $\epsilon$ follows Eq. \eqref{eq:FI-Gaussian-app}.
Utilizing Eq.\eqref{eq:AverageResults-app}, we obtain
\begin{equation}
	F^{\epsilon}=\int \frac{d \nu}{2\pi} \left[4\frac{|\alpha_{\nu}|^{2}}{2\bar{n}^{\prime}_{\nu}+1} \left|\frac{d S_{\nu}}{d \epsilon}\right|^{2}+\frac{|\partial_{\epsilon}\bar{n}^{\prime}_{\nu}|^{2}}{\bar{n}^{\prime}_{\nu}(\bar{n}^{\prime}_{\nu}+1)}\right],
	\label{eq:QFI-app}
\end{equation}
where $S_{\nu}=\gamma_{ex}G_{a}(\nu)$ characterizes the scattering amplitude. The explicit expressions of $\bar{n}^{\prime}_{\nu}$ and $S_{\nu}$ are obtained as
\begin{eqnarray}
	\bar{n}^{\prime}_{\nu}&=&  \bar{n}_{\nu}+\frac{\gamma_{ex} \kappa g^{2} S_{z} g^{2}_{G} (\bar{n}_{\nu}+\frac{N_{e}}{S_{z}})}{|(\nu-\nu_{1})(\nu-\nu_{2})(\nu-\nu_{3})|^{2}},\label{eq:Snu-app}\\
	S_{\nu}&=& \gamma_{ex}\frac{(\nu-\nu_{b}+i \frac{\gamma_{b}}{2})(\nu-\nu_{b}+i \frac{\kappa}{2})+S_{z} g^{2}_{G}}{(\nu-\nu_{1})(\nu-\nu_{2})(\nu-\nu_{3})}.
	\label{eq:nT-app}
\end{eqnarray}
Here $\nu_{i=1,2,3}$ are the eigenvalues of the coefficient matrix
\begin{equation}
	M=\nu_{b} \mathbb{I}+
\left(
\begin{array}{ccc}
	\epsilon -i \frac{\gamma_{a}^{\prime}}{2}& g &0\\
	g&-i\frac{\gamma_{b}}{2} & \sqrt{S_{z}} g_{G}\\
	0&-\sqrt{S_{z}} g_{G}&-i\frac{\kappa}{2}\\
\end{array}
\right).
\label{eq:CoeMatrix-app}
\end{equation}
Comparing Eqs. (\ref{eq:Snu-app}) and (\ref{eq:nT-app}) with \eqref{eq:CoeMatrix-app}, we find that both $S_{\nu}$ and $\bar{n}^{\prime}_{\nu}$ are well defined unless the matrix $M_{\nu}\equiv (\nu \mathbb{I} -M)$ is singular, i.e., $\mathrm{det}[M_{\nu}]=0$. This singular condition indicates a lasing transition. Below the lasing threshold, $\partial_{\epsilon} S_{\nu}$ and $\partial_{\epsilon} \bar{n}_{\nu}^{\prime}$ are well defined, so the QFI shows no singularity. Near the singular point, we have $\det[M_{\nu}]\approx 0$, $\bar{n}_{\nu}\sim |M^{-1}_{\nu}|^{2}$, $S_{\nu}\sim |M_{\nu}^{-1}|$. Thus the QFI scale as
\begin{equation}
	F^{\epsilon} \sim \int \frac{d \nu}{2 \pi} |M^{-1}_{\nu}|^{2}.
	\label{eq:ScalingQFI-app}
\end{equation}
The validity of Eq. \eqref{eq:ScalingQFI-app} is based on the linear description of the sensing system. The linearization is invalid near the transition point where the critical fluctuations diverge. The diverging critical fluctuations may prevent the QFI from diverging.

For $\kappa\gg \gamma_{a}$ and $\gamma_{b}$, one can adiabatically eliminate the gain medium mode and get an effective two dimensional coefficient matrix
\begin{equation}
	M_{\mathrm{eff}}=\nu_{b} \mathbb{I}
+\left(
\begin{array}{cc}
	\epsilon-\frac{i \gamma^{\prime}_{a}}{2} & g \\
	g& -i\frac{\gamma_{b}^{\prime}}{2}\\
\end{array}
\right),
\label{eq:CoeMatrixEff-app}
\end{equation}
where $\gamma_{b}^{\prime}=\gamma_{b}-\frac{4{S_{z}} g_{G}^{2}}{\kappa}$ is the effective decay rate of cavity $b$. The EP occurs at $\epsilon=0$ and $g=\frac{1}{4}(\gamma^{\prime}_{a}-\gamma^{\prime}_{b})$. The lasing threshold is determined by the singular condition $\mathrm{det}[\nu \mathbb{I}-M_{\mathrm{eff}}]=0$, which yields
\begin{equation}
	S_{c}=\left\{
	\begin{array}{ll}
		\frac{\kappa}{4 g_{G}^{2}}(\frac{4 g^{2}}{\gamma^{\prime}_{a}}+\gamma_{b}), & g<\frac{\gamma^{\prime}_{a}}{2};\\
		\frac{\kappa}{4 g_{G}^{2}}(\gamma^{\prime}_{a}+\gamma_{b}), & g\ge\frac{\gamma^{\prime}_{a}}{2}.
	\end{array}
	\right.
	\label{eq:Threshold-app}
\end{equation}
Here we suppose the cavities are in resonance so that $\epsilon=0$. 
By implementing the same approximation to the gain medium propagators, we obtain $G_{d}(\nu)\approx G_{d}^{(0)}(\nu)\approx \frac{2 S_{z} g^{2}_G}{\kappa}$, and hence
\begin{eqnarray}
	S_{\nu}&\approx& \gamma_{ex} \frac{(\nu-\nu_{b}+i \frac{\gamma^{\prime}_{b}}{2})}{(\nu-\nu_{+})(\nu-\nu_{-})}, \nonumber \\
	\bar{n}^{\prime}_{\nu}&\approx& \bar{n}_{\nu}+\frac{\gamma_{ex} g^{2}\frac{4 S_{z} g^{2}_{G}}{\kappa} (\bar{n}_{\nu}+\frac{N_{e}}{S_{z}})}{|(\nu-\nu_{+})(\nu-\nu_{-})|^{2}},
	\label{eq:SandTApp-app}
\end{eqnarray}
where $\nu_{\pm}$ are the eigenvalues of $M_{\mathrm{eff}}$.  Applying these results into Eq. \eqref{eq:QFI-app}, the QFI of the active-passive coupled cavity system is obtained. Around the threshold, the same conclusion as Eq. \eqref{eq:ScalingQFI-app} is obtained only with the substitution of $M_{\mathrm{eff}}$ for $M$.

\bibliographystyle{apsrev4-1}
\bibliography{NHM}
\end{document}